\documentstyle[aps,multicol,graphics]{revtex}

\newcommand{\ket}[1]{\left| #1 \right\rangle}
\newcommand{\bra}[1]{\left\langle #1\right |}
\begin{document}
\draft
\title{Quantum Search with Two-atom Collisions in Cavity QED}
\author{F. Yamaguchi$^1$\thanks{Electronic address: yamaguchi@stanford.edu},
        P. Milman$^2$, M. Brune$^2$, J. M. Raimond$^2$, S. Haroche$^{2,3}$}
\address{$^1$ E. L. Ginzton Laboratory, Stanford University, Stanford, CA 94305, USA}
\address{$^2$ Laboratoire Kastler Brossel,
            D\'epartement de Physique de l'Ecole Normale Sup\'erieure, \\
            24 rue Lhomond, F-75231 Paris Cedex 05 France}
\address{$^3$ Coll\`ege de France, 11 place Marcelin-Berthelot, F-75005,
            Paris France}
\date{\today}
\maketitle
\begin{abstract}
We propose a scheme to implement two-qubit Grover's quantum search algorithm
using Cavity Quantum Electrodynamics. Circular Rydberg atoms are used as quantum bits (qubits).
They interact with the electromagnetic field of
a non-resonant cavity . The quantum gate dynamics
is provided by a cavity-assisted collision, robust against decoherence
processes. We present the detailed procedure and analyze the experimental feasibility.
\end{abstract}

\begin{multicols}{2}

\narrowtext

Quantum mechanics makes it possible in principle to realize new
information processing functions \cite{Deutsch:R.Soc.Lond1985}, ranging
from the relatively simple quantum cryptography protocols \cite{BB84} to complex 
quantum calculation algorithms \cite{Shor:SIAM1997,Grover:PRL1997}. They are based
on manipulations of quantum entanglement between two-level quantum systems
or qubits. 

The practical implementation of quantum information processing,
requiring an excellent isolation of the qubits from the environment,
puts very severe constraints on the experimental systems. Many
efforts have been recently devoted to the evaluation of various experimental
approaches to quantum bits and quantum gates: trapped ions \cite{Wine:PRL1995}, cavity
quantum electrodynamics \cite{Rauschenbeutel:PRL1999}, liquid state NMR
\cite{Gershenfeld:Science1997}, superconducting mesocircuits \cite{esteve}. They culminated
in the implementation of simple quantum algorithms \cite{Gershenfeld:Science1997}.

In this context, cavity QED with circular Rydberg atoms and superconducting cavities
presents a peculiar interest. The qubits are carried by long lived atomic levels
or cavity states. Both the initial and final states of these qubits can be determined precisely.
The resonant atom-cavity interaction, resulting in an energy
exchange between the atom and the field, provides a direct mechanism to entangle
the atomic and the cavity states \cite{Raimond:RMP2001} or to realize an atom-cavity 
quantum gate \cite{Rauschenbeutel:PRL1999}. Using successive
interactions of a series of atoms with the same cavity mode, we have tailored various entangled
states such as EPR pairs \cite{Hagley:PRL1997} and GHZ triplets of entangled particles 
\cite{Rauschenbeutel:Science2000}. In these experiments,
the quantum information is transiently stored in the cavity mode. The final fidelity is thus
limited by the cavity losses, which are the main cause of decoherence.

We have recently demonstrated an alternative approach to quantum entanglement generation in cavity
QED \cite{Zheng:PRL2000,Osnaghi:PRL2001}. Two atoms directly interact with each other 
through a van der Waals interaction, assisted by
the cavity mode. The entanglement dynamics only involves the virtual exchange of
a photon with the field. To first order, the scheme is insensitive to cavity losses or to
the presence of a stray thermal field in the mode. This new type of quantum gate opens interesting
perspectives for quantum information processing in the cavity QED context. We show here that a simple
extension of this experiment can be used to implement the two-qubit Grover search algorithm, with a
high fidelity. This is the first proposal, to our knowledge, of implementation of this search algorithm 
in CQED. Note that cavity QED implementation of another algorithm has been independently proposed
\cite{SCULLY}

Let us recall briefly the main features of Grover's search algorithm \cite{Grover:PRL1997}.
The goal is to find one item among $N$, which are stored in an unsorted
database. The database can be accessed by an ``oracle", a ``Black
box" comparing any item with the searched one. It gives the ``Yes" answer
when the items match, ``No" overwise.
The most efficient classical algorithm is to examine items one by
one until the blackbox returns ``Yes". On average, $N/2$
inquires are necessary. In
the quantum search algorithm\cite{Grover:PRL1997}, multiple items
are simultaneously examined using a superposition of the
corresponding states. The quantum search for the marked item requires only $O(\sqrt{N})$
inquires.

More precisely, the database items are represented by a quantum register with
$n$ qubits, having $N=2^n$ possible states, $\ket{0}=\ket{000\cdots
0}$, $\ket{1}=\ket{000\cdots 1}$, $\cdots$,
$\ket{2^n-1}=\ket{111\cdots 1}$. Let us assume that the marked item corresponds to state
$\ket{\tau}$ and that the ``Yes"/``No" answer from this blackbox
is coded in a $\pi/0$ phase shift. The transformation performed by the oracle is thus 
$I_{\tau}=I-2\ket{\tau}\bra{\tau}$, where $I$ is the $2^n\times 2^n$ identity matrix on 
the quantum register. Note that $I_\tau$ amounts to a conditional phase operation.
 The algorithm consists in a repetition of the
transformation:  
\begin{equation}
  Q\equiv HI_0 H I_{\tau} \label{eq:Grover2}\ ,
\end{equation}
(time proceeds from right to left), where $I_0=I-2\ket{0}\bra{0}$ and $H=\prod_i^n H_i$ is 
the product of  Hadamard gates acting on the $i$-th qubit. $H_i$ transforms each qubit as   
\begin{equation}
    H_i:\left\{\matrix{
    \ket{0}_i\rightarrow \frac{1}{\sqrt{2}} (\ket{0}_i+\ket{1}_i)\cr
    \ket{1}_i\rightarrow \frac{1}{\sqrt{2}} (\ket{0}_i-\ket{1}_i)}
    \right. 
	\end{equation}

The sequence of $Q$ operations acts on a state prepared initially in $\ket{\Psi}=\prod_i^n \ket{\psi}_i$, 
where $\ket{\psi}_i= H_i \ket{0}_i$. 
The initial state $\ket{\Psi}$ is a superposition
of all computational states with equal amplitudes,
$\frac{1}{\sqrt{N}}\sum_{i=1}^N\ket{i}$. 
All states are then
examined simultaneously by the blackbox, $I_{\tau}$. Only state
$\ket{\tau}$ gains minus sign and the resulting register state is
$\frac{1}{\sqrt{N}}\left(\sum_{i\neq\tau}\ket{i}-\ket{\tau}\right)$. The
operators $HI_0H$ finally perform an ``inversion about the average"
operation, which increases the probablity amplitude of $\ket\tau$.
After $O(\sqrt{N})$ iterations of this elementary transformation, the probability to
get the register in state $\ket\tau$ is maximum and of the order of unity. A simple
read-out of the register provides thus the searched item.

In the simple case case of two qubits ($n=2$), on which we will focus
from now on, there are only $4$ items,
$\ket{0}=\ket{00}$, $\ket{1}=\ket{01}$, $\ket{2}=\ket{10}$ and
$\ket{3}=\ket{11}$. After the first oracle operation, 
the average of the amplitudes of the four states,
$\frac{1}{2}$ for $i\neq\tau$ and $-\frac{1}{2}$ for $i=\tau$, is
$\frac{1}{4}$, and the inversion about $\frac{1}{4}$ leads to the
amplitudes $0$ for $i\neq\tau$ and 1 for $i=\tau$. 
Grover's search thus requires only one tranformation $Q$ and 
one inquiry of the blackbox (note that the classical search requires 
in this case two inquiries on the average). 

The $H_i$ transformations are single qubits gates, easily performed in any
physical implementation. The most critical part of the algorithm is the realization
of the $I_\tau$ and $I_0$ transformations, which produce an entangled state of the two
qubits. We show now that the Grover operation $Q$ reduces to single qubit gates and to two
applications of the quantum phase gate $I_{QPG}$, defined by the unitary matrix:
\begin{equation}
    \pmatrix{
            1 & 0 & 0 & 0\cr
            0 & 1 & 0 & 0\cr
            0 & 0 & 1 & 0\cr
            0 & 0 & 0 & -1
         }
    \equiv I_{\rm QPG}
\end{equation}

The operators $I_{\tau}$ can be obtained by adding rotations of qubits 1 and 2 about
$z$-axis of $\theta_1$ and $\theta_2$, respectively, to $I_{\rm
QPG}$:
\begin{eqnarray}
    &&Z_1(\theta_1)Z_2(\theta_2)I_{\rm QPG}
    =I_{\rm QPG}Z_1(\theta_1)Z_2(\theta_2)\nonumber\\
    &&=\pmatrix{
            e^{-\frac{i}{2}(\theta_1+\theta_2)} & 0 & 0 & 0\cr
            0 & e^{-\frac{i}{2}(\theta_1-\theta_2)} & 0 & 0\cr
            0 & 0 & e^{\frac{i}{2}(\theta_1-\theta_2)} & 0\cr
            0 & 0 & 0 & -e^{\frac{i}{2}(\theta_1+\theta_2)}
        }\ .
    \label{eq:Z-QPG}
\end{eqnarray}
The relevant rotation angles $\theta_1$ and  $\theta_2$ are $(\pi, \pi)$, 
$(0, \pi)$, $(\pi, 0)$ and $(0,0)$ for implementing
$I_{\tau}$ up to a global phase ($\ket{\tau}=\ket{00}$, $\ket{01}$, $\ket{10}$, and $\ket{11}$,
respectively). 

The sequence described by Eq.(\ref{eq:Grover2}) for 2-bit Grover's
algorithm can be decomposed as
\begin{equation}
    \left[H_1 Z_1(\pi)\right]
    \left[H_2Z_2(\pi)\right]
    I_{\rm QPG} H I_{\rm QPG}\left[Z_1(\theta_1)H_1\right]
    \left[Z_2(\theta_2)H_2\right]\ .
\end{equation}

It can be further simplified by the relationship,
\begin{equation}
    Z_i(\pm\theta)=H_i X_i(\mp \theta)H_i,
\end{equation}
where $X_i(\theta)$ corresponds to the unitary transformation,
\begin{equation}
    \matrix{
        \ket{0_j}\rightarrow
            \cos\frac{\theta}{2}\ket{0_j}
          +i\sin\frac{\theta}{2}\ket{1_j},\cr
        \ket{1_j}\rightarrow
            i\sin\frac{\theta}{2}\ket{0_j}
            +\cos\frac{\theta}{2}\ket{1_j}\ .
    }
\end{equation}

Using $H_i^2=1$, the final transformation performing the whole Grover search writes thus:
\begin{equation}
    S I_{\rm QPG} H
    I_{\rm QPG} P.
    \label{eq:Grover2_sequence}
\end{equation}
In Eq.(\ref{eq:Grover2_sequence}), $S$ and $P$ are abbreviations
for the operations $S_1 S_2$ and $P_1(\theta_1) P_2(\theta_2)$,
respectively. The operation $S_j$ is defined as
$X_i(-\pi)H_j$ and writes explicitely,
\begin{equation}
    S_j:\left\{
        \matrix{
            \ket{0}\rightarrow\frac{i}{\sqrt{2}}(-\ket{0}-\ket{1})\cr
            \ket{1}\rightarrow\frac{i}{\sqrt{2}}(\ket{0}-\ket{1})
        }\right.\ .
    \label{eq:R}
\end{equation}
The transformations $P_j(\theta)\equiv
H_jX_j(-\theta)$ can be written also in a compact form as
\begin{eqnarray}
    &&P_j(\theta):\left\{
        \matrix{
            \ket{0_j}\rightarrow\frac{1}{\sqrt{2}}(e^{-\frac{i\theta}{2}}\ket{0_j}+e^{\frac{i\theta}{2}}\ket{1_j})\cr
            \ket{1_j}\rightarrow\frac{1}{\sqrt{2}}(e^{-\frac{i\theta}{2}}\ket{0_j}-e^{\frac{i\theta}{2}}\ket{1_j})
        } \right.
    \label{eq:P}
\end{eqnarray}
We have thus finally expressed the Grover search as a simple sequence of
single qubits rotations interrupted by two quantum phase gate operations only.

The scheme of the proposed implementation is displayed on figure \ref{fig:cavity}.
The cavity $C$ is a Fabry-Perot resonator sustaining a resonant mode with a gaussian
transverse geometry and a standing wave pattern along the cavity axis. 
Two atomic beams effusing from the same oven $O$ cross the superconducting millimeter-
wave cavity $C$ at two separate antinodes in the mode standing wave. Two atoms
$A_1$ and $A_2$, carrying qubits 1 and 2, are simultaneously prepared in box $B$ 
in each atomic beam into a high lying circular
Rydberg state. The relevant atomic levels
($\ket{e_j}$, $\ket{g_j}$, and $\ket{i_j}$) are shown in
Fig.\ref{fig:cavity}. The logical states $1$ and $0$ of qubit 1 are represented by
$\ket{e_1}$ and $\ket{g_1}$ states of A$_1$ respectively. 
Qubit 2 uses instead as logical levels 1 and 0 states $\ket{i_2}$ 
and $\ket{g_2}$ of A$_2$ respectively (this choice is imposed by the quantum phase gate operation).
Both atoms have the same velocity $v$. They interact together with the cavity mode and are finally
detected separately in the state-selective field ionization detectors $D_1$ and $D_2$. 

While they cross the cavity mode, the atoms undergo single qubits rotations 
(transformations $P$, $H$ and $S$). They are produced by external classical microwave sources
resonant on the $e\rightarrow g$ transition for $A_1$ and $g\rightarrow i$
transition for $A_2$. The amplitude and phase of these sources are carefully tuned to 
produce the required transformations. It is important that the microwave used for atom $A_1$ does not affect
atom $A_2$ by mixing the $g$ and $e$ levels. In the same way, the microwave used for $A_2$ must not interact
with $A_1$. It is thus essential that the atomic transitions have slightly different
frequencies for the two atoms. We plan to use a set of electrodes creating in the cavity
an inhomogeneous electric field, used to tune the atomic transitions through the Stark effect. Since the two atoms
experience different fields, their frequencies can be controlled independently. Separate interactions with the two
classical microwave sources can then be tailored.

In between these single qubit rotations, the atoms experience two
dispersive interactions with the cavity mode providing, through the cavity-assisted van der Waals collision, the
quantum phase gate $I_{QPG}$ dynamics. For these interactions, the inhomogeneous component of the cavity field is 
suppressed. A voltage applied across the cavity mirrors provide an homogeneous field used to tune the common atomic
frequencies at the proper value. 

Let us now describe in more details the cavity assisted
van der Waals collision between $A_1$ and $A_2$. 
The cavity field is assumed to be in its
vacuum  state at the beginning of the two-atom interaction. In a dispersive regime,
where the detuning $\delta$ between the $e\rightarrow g$
atomic transition frequency $\omega_0$ and the cavity frequency $\omega$ is much
greater than the atom-cavity coupling $\Omega$
($\delta\gg\Omega$), the Hamiltonian for the two-atom
system can be approximated by the effective expression \cite{Zheng:PRL2000}
\begin{equation}\label{heff}
    H_{eff}=\lambda\left[\sum_{j=1,2}\ket{e_j}\bra{e_j}
    +(S_1^{+}S_2^{-}+S_1^{-}S_2^{+})\right],
\end{equation}
where $\lambda=\Omega^2/4\delta$, $S_j^{+}=\ket{e_j}\bra{g_j}$ and
$S_j^{-}=\ket{g_j}\bra{e_j}$.
The first sum in $H_{eff}$ describes the cavity Lamb shift (or
vacuum light shift) experienced by the two atoms. The second term describes the
energy exchange between the atoms mediated by the cavity field.

When the interaction time $t$ is chosen so that $\lambda t=\pi$,
the two-atom system undergoes the transition,
\begin{equation}
    \matrix{
        \ket{g_1}\ket{g_2}\rightarrow \ket{g_1}\ket{g_2}, \cr
        \ket{g_1}\ket{i_2}\rightarrow \ket{g_1}\ket{i_2}, \cr
        \ket{e_1}\ket{g_2}\rightarrow \ket{e_1}\ket{g_2}, \cr
        \ket{e_1}\ket{i_2}\rightarrow -\ket{e_1}\ket{i_2}.
  	}
\end{equation}
This transformation corresponds to a conditional quantum phase
gate (QPG) between the two qubits.

Let us discuss now the practical feasibility of this experiment. 
The coupling of the atoms to the cavity field is
$\Omega/2\pi$ = 50kHz\cite{Osnaghi:PRL2001}. In order to get a good
entanglement in the cavity enhanced collision, the detuning $\delta$ should
be much larger than $\Omega$. We choose here 
$\delta/2\pi=4\Omega$. This setting matches the dispersive regime requirement 
and provides a gate operation well described
by $H_{eff}$. With this setting, the atom-cavity interaction time should be
$2.5\times 10^{-4}$ s to allow for two QPG gates operation. The time needed
for the single qubit rotations is negligible at this scale. The velocity of the atoms should
thus be of the order of 40 m/s. This value is within reach of simple atomic beam techniques
with transverse laser cooling to increase the density of slow atoms. The total interaction time 
with the mode is thus 120 $\mu$s, short compared to the photon lifetime, 
1 ms in the present cavity \cite{Raimond:RMP2001}. 

We have performed a numerical simulation of the experiment to estimate the achievable fidelity. 
We used for that the exact hamiltonian describing the non resonant atom field coupling, in order 
to test the validity of the effective hamiltonian (\ref{heff}) approximation.  Since the field 
in the cavity is only virtually populated, dissipation effects were not considered.  

As is shown in Fig.~\ref{fig:simu}(a) an efficiency of $\approx 94\%$ is achieved. The limited 
fidelity is due to that the real hamiltonian does not exactly produce the $I_{QPG}$ realized
by its approximation $H_{eff}$. We also considered in the simulation the possibility of 
imperfection in the classical pulses duration. Their role is to decrease the probability of 
obtaining the searched state ate the end of the proccess. By considering pulse imperfections 
of the order of $5\%$ in the
simulations, we observed that the efficiency of our search decreased to $85\%$, as shown
in Fig.~\ref{fig:simu} (b). The dependency of the efficiency of our experimental proposal
on pulse imperfections can be seen in Fig.~\ref{fig:simu}(c), where the fidelity of the
final state is plotted against the error in the classical pulses applied in the atoms.

We proposed here a simple implementation of the two qubit Grover search algorithm. It can be 
realized with minor amendments of our Rydberg atom-cavity setup. It explicitely makes use of
qubit entanglement. Its experimental implementation will be an illustration of the power of
cavity QED to manipulate complex entangled states for quantum information processing.

This work was supported by Japan Science and Technology
Corporation and European Community (International Cooperative
Research Project, ``Quantum Entanglement"). Laboratoire Kastler
Brossel, Universit\'{e} Pierre et Marie Curie and ENS, is
associated with CNRS (UMR 8552).

%%%%%%%%%%%%%%%%%%%%%%%%%%%%%%%%%%%%%%%%%%%%%%%%%%%%%%%%%%%%%%%%%%%
% References
%%%%%%%%%%%%%%%%%%%%%%%%%%%%%%%%%%%%%%%%%%%%%%%%%%%%%%%%%%%%%%%%%%%

\end{multicols}

\vspace*{5cm}

\begin{figure}
\begin{center}
%\special{eps:fig1.eps x=9cm y=5cm}
	
\end{center}
\caption{Experimental apparatus. Atoms $A_1$ and $A_2$ cross the cavity with same velocity $v$ but at different positions, 
allowing for individual manipulation of each one in the regions corresponding to pulses $H$, $S$ and $P$. The phase gates ($I_{QPG}$) are applied between each pair of classical pulse as shown in the figure. In the inset, the atomic level scheme with the corresponding frequencies.} \label{fig:cavity}
\end{figure}

\begin{figure}
\begin{center}
%\scalebox{1.}{\includegraphics{fig2.eps}}
\end{center}
\caption{Results of the numerical simulation of the proposed experiment (a) Fidelity for the case where the item searched corresponds to state $\ket{ei}$. The state is obtained with a probability of $\approx 94\%$. (b) Fidelity when imperfections in the classical pulse are considered. The resulting state, for the case in which there is a $5\%$ error in the classical pulses duration, coincides with the desired one with an $\approx 85\%$ probability. (c) Dependance of the fidelity on the pulse imperfections. In real experiments this imperfection can be of the order of $3\%$.} \label{fig:simu}
\end{figure}

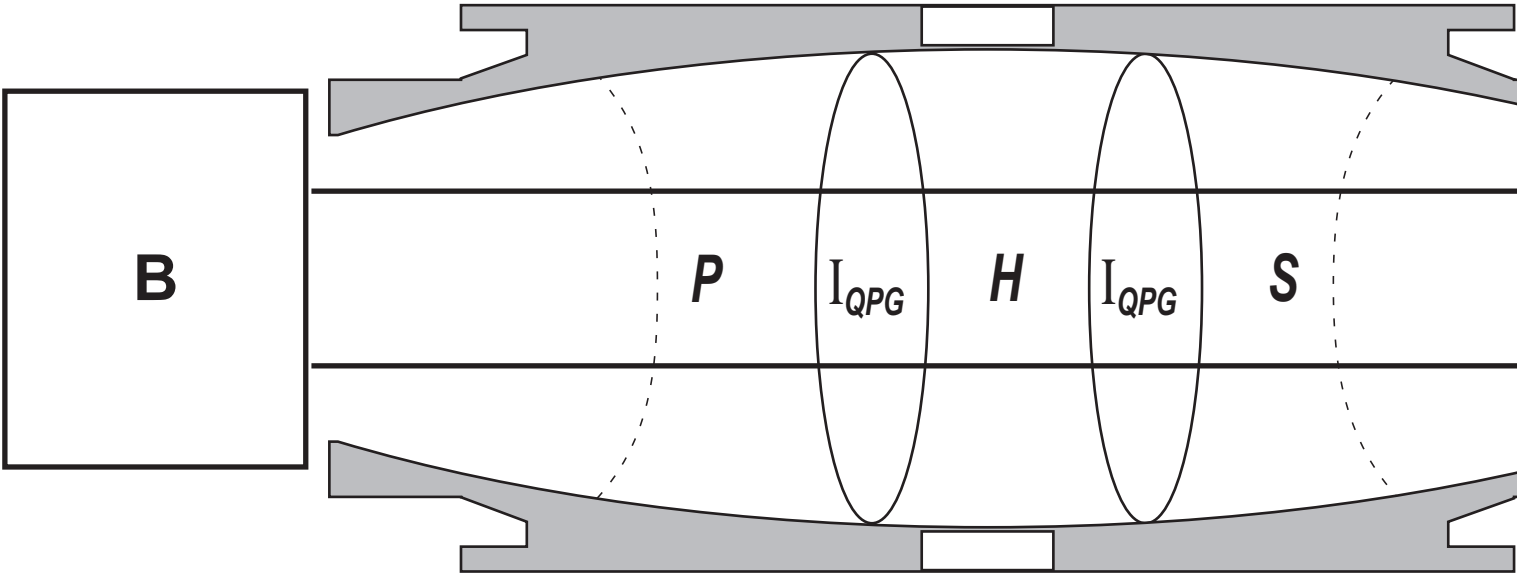
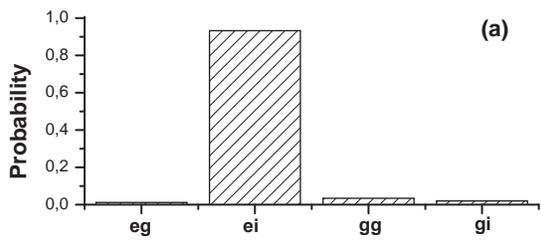
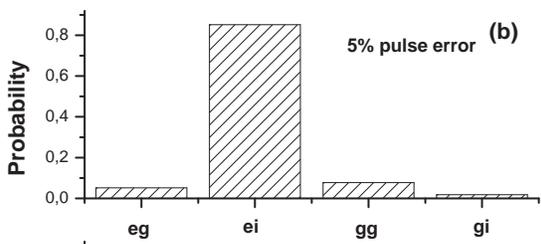
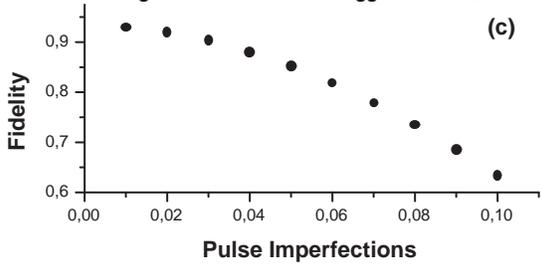
\end{document}